\documentclass[aps,pra,groupedaddress]{revtex4}

\usepackage{graphicx}
\usepackage{dcolumn}
\usepackage{bm}
\usepackage{color}

\begin{document}


\title{Off-resonant Raman transitions impact in an atom interferometer}

\author{A. Gauguet}
\author{T. E. Mehlst\"{a}ubler\footnote{Present address: Physikalisch-Technische Bundesanstalt, Bundesallee 100, D-38116 Braunschweig, Germany}}
\author{T. L\'{e}v\`{e}que}
\author{J. Le Gou\"{e}t\footnote{Present address: Optical and Quantum Communications Group, MIT, Rm 36-479,Cambridge, MA 02139, USA}}
\author{W. Chaibi}
\author{B. Canuel\footnote{Present address: European Gravitational Observatory, Via E. Amaldi, 56021 S. Stefano a Macerata - Cascina (PI), Italy}}
\author{A. Clairon}
\author{F. Pereira Dos Santos}
\author{A. Landragin}

\email[]{arnaud.landragin@obspm.fr}
\affiliation{LNE-SYRTE, UMR 8630 CNRS, Observatoire de Paris, UPMC, 61 avenue de l'Observatoire, 75014 Paris,
FRANCE}

\date{\today}

\begin{abstract}
We study the influence of off-resonant two photon transitions on high precision measurements with atom interferometers based on stimulated Raman transitions. These resonances induce a two photon light shift on the resonant Raman condition. The impact of this effect is investigated in two highly sensitive experiments: a gravimeter and a gyroscope-accelerometer. We show that it can lead to significant systematic phase shifts, which have to be taken into account in order to achieve best performances in term of accuracy and stability.
\end{abstract}


\maketitle
\section{INTRODUCTION}

In the field of atom interferometry, the improving sensitivity of inertial sensors~\cite{Ackim,gyro,GyroPRL,gradio} is paving the way for many new applications in geophysics, navigation and
tests of fundamental physics. Most of these experiments are based on Raman transitions~\cite{Chu91} to realize beamsplitters and mirrors, which manipulate
the atomic wave-packets. Among others, this technique has the advantage of an internal state labelling of the exit ports of the interferometer~\cite{borde}, enable an
efficient detection methods. Moreover, the atoms spend most of the time in free fall, with very small and calculable interactions with the environment. The inertial forces are then determined by the relative
displacement of the atomic sample with respect to the equiphases of the laser beams, which realise a very accurate and stable ruler. This makes this technique suitable for high precision measurements, as required for instance for inertial sensors and for the determination of fundamental constants~\cite{Gf,Gl,alphaW,alphaC,BW}.

A limit to the accuracy and the long term stability of these sensors comes from wave-front distortions of the
laser beams. This wave-front distortion shift appears
directly on the signal of an interferometer when the atoms experience different wave-fronts at each Raman pulse. This effect thus depends on the actual trajectories of the atoms, so that a precise
control of the initial position, velocity and temperature of the atomic clouds is required~\cite{jerome,varenna}.  A convenient 
technique to reduce this bias is to minimize the number of optical components in the shaping of the two Raman laser beams and by implementing them in a retro-reflected
geometry~\cite{Ackim,GyroPRL,GraviAPB1}. Indeed, as long as the two beams travel together, wave-front aberrations are identical for the two beams and thus have no influence on their phase difference. This geometry also provides an efficient way to use the \textbf{k} reversal
technique, which allows to diffract the atomic wavepackets in one or the opposite direction and thus to separate effects of many major
systematic errors such as gradients of magnetic fields or light shifts~\cite{weiss94}. The main drawback of this geometry
arises from the presence of off-resonant Raman transitions, which induce a light shift on the resonant Raman transition and thus a phase shift of the atom interferometer.

In the following, we investigate this effect called two photon light shift
(TPLS)~\cite{Biraben2006}. We first show that the TPLS arises from several off-resonant transitions and evaluate each contribution. We then derive the impact onto the phase of an atom interferometer and use our gravimeter and gyroscope-accelerometer for quantitative comparisons. In particular we measure the systematic shifts and we investigate the influence on the long term stability.
 The study demonstrates that the precise control of experimental parameters, in particular the Raman laser intensities and polarisations, is needed to reduce the influence of this effect for such interferometers.

\section{Light Shift due to off-resonant Raman transitions}

\subsection{Raman Spectroscopy}

\begin{figure}[!h]
 \includegraphics[width=8.5cm]{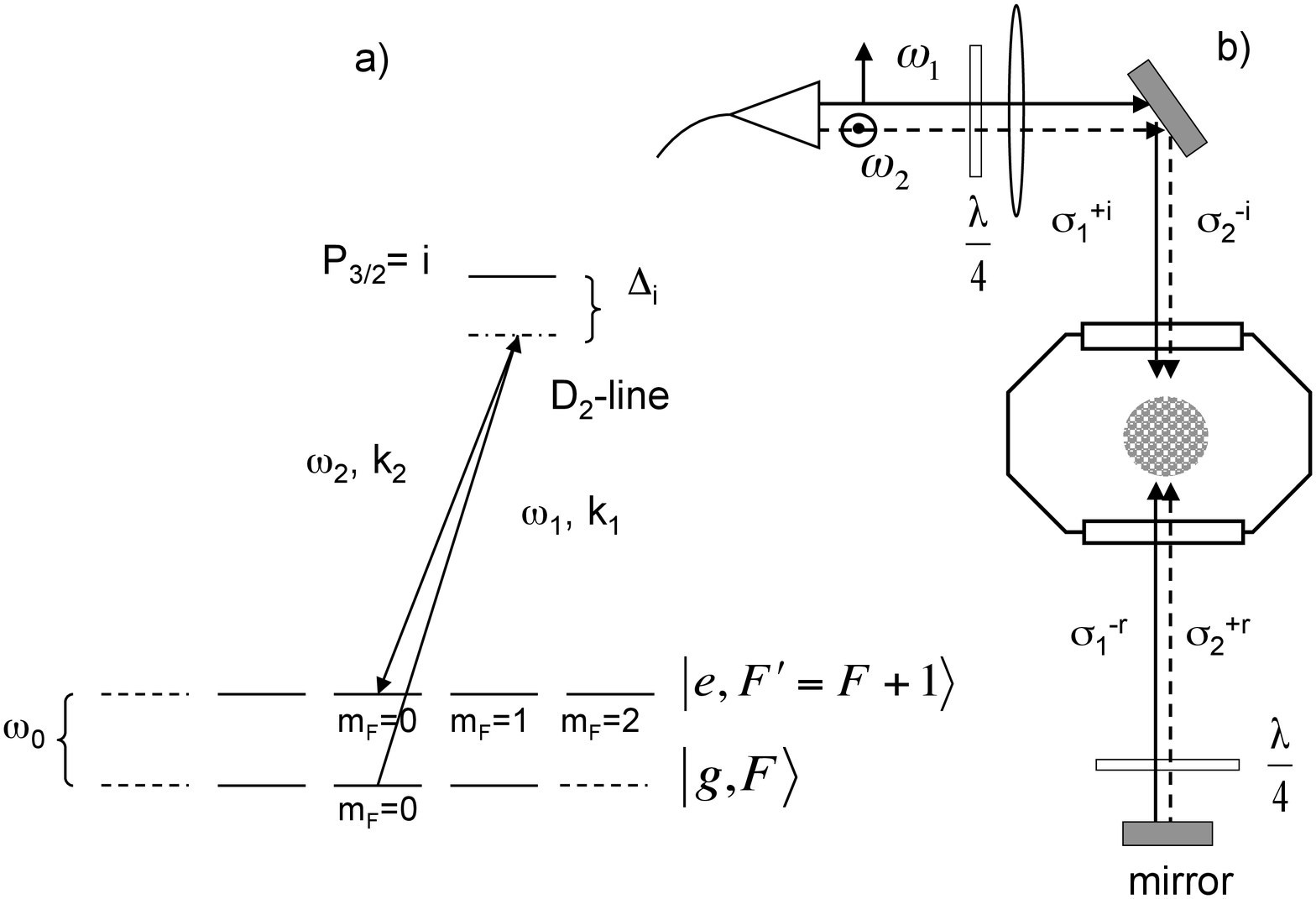}
 \caption{Scheme of our Raman laser setup. a) Scheme of the Raman transition between the two hyperfine ground states of an alkaline atom. b) Implementation of the Raman laser beams. The two Raman lasers of orthogonal polarisations are guided to the experiment through the same polarization maintaining fibre. The two lasers are represented respectively by the solid and dashed lines. The lasers go through a first quater-wave plate ($\lambda/4$), cross the experiment and are reflected by a mirror, crossing twice a second quater-wave plate. The wave-plates are set in such a way that counter-propagating Raman transitions are allowed but co-propagating Raman transitions are forbidden.}
 \label{Schema}
 \end{figure}

The two experiments are using different alkali-metal atoms: $^{87}$Rb in the case of the gravimeter and $^{133}$Cs in the case of the gyroscope.  As hyperfine structures, transition selection rules and Raman laser setups are similar (see figure~\ref{Schema}), their results can be compared easily. The Raman transitions couple the two hyperfine ground states of the alkaline atom (labelled $|g\rangle$ and $|e\rangle$) via an intermediate state (labelled $|i\rangle$) and two lasers with frequencies (labelled  $\omega_1$ and  $\omega_2$) detuned by $\Delta_i$ on the red of the $D_{2}$ line.  During the interferometer sequence, a bias magnetic field is applied along the direction of propagation of the Raman laser beam to lift the degeneracy of the magnetic sublevel manifold. 
The two Raman lasers are overlapped with orthogonal linear polarisations and delivered within the same polarisation maintaining optical fiber to the vacuum chamber. After the fiber,
the Raman beams pass through a quarter-wave plate to convert the initial linear polarisations into circular polarisations, noted $\sigma_1^{i+}$ for the Raman laser at the frequency $\omega_1$
and $\sigma_2^{i-}$ for the orthogonal polarisation at $\omega_2$. These beams are then retro-reflected through a quarter-wave plate to rotate the polarisation of each beam into its orthogonal polarisation ($\sigma^{+r}_2$, $\sigma^{-r}_1$).

For $m_F=0$ to $m_F=0$ transitions, there are two pairs of beams ($\sigma^{+}/\sigma^{+}$ and $\sigma^{-}/\sigma^{-}$), which can drive counter-propagating Raman transitions with
effective wave-vectors $\pm\mathbf{k}_\mathrm{eff} = \pm(\mathbf{k}_1-\mathbf{k}_2)$. Then, the ground state $|g,\mathbf{p}\rangle$ is coupled with the excited state $|e,\mathbf{p}+\hbar
\mathbf{k}_\mathrm{eff}\rangle$ by the pair of Raman laser ($\sigma^{+i}_1 / \sigma^{+r}_2$) and to the excited state $|e,\mathbf{p}-\hbar \mathbf{k}_\mathrm{eff}\rangle$ with the pair of
Raman laser ($\sigma^{-i}_2 / \sigma^{-r}_1$). 

We use the Doppler effect to lift the degeneracy between the two resonance conditions. Indeed, if the atoms have a velocity in the direction of propagation of the Raman lasers, the Doppler
shifts are of opposite sign for the two counter-propagating transitions. The resonance condition for each of these couplings is $\Delta \omega_\mathrm{laser} = \omega_{0} +
\omega_\mathrm{r}\pm \omega_\mathrm{D}$, where $\omega_{0}$ is the hyperfine transition frequency, $\hbar \omega_\mathrm{r} = \hbar k_\mathrm{eff}^2/2m$ is the
recoil energy and $\omega_\mathrm{D} =-\mathbf{k}_\mathrm{eff}.\mathbf{v}$ the Doppler shift due to the atomic velocity $\mathbf{v}$ in the reference frame of the
apparatus. Consequently, the detuning between the two resonances is $2 \omega_\mathrm{D}$, therefore we can discriminate between the two transitions when the Doppler shift is large enough compared to
the linewidth of the Raman transition. This linewidth is characterised by the effective Rabi frequency  $\Omega_\mathrm{eff}$, which depends on the product of the two Raman lasers intensities and inversely to the Raman detuning $\Delta_i$~\cite{weiss94}. 

In this first part, we use the gyroscope-accelerometer experiment described in detail in ~\cite{GyroPRL}. The experiment has been performed with a cloud of cold Caesium atoms  (1.2 $\mu K$)
prepared initially in the $|F=3,m_F=0\rangle$ state. The atoms are launched at 2.4 m.s$^{-1}$ with an angle of $8^\circ$ with respect to the vertical direction. The Raman lasers are implemented in the
horizontal plane with a $6^\circ$ angle with the normal of the atomic flight direction. Thus, to select only one Raman transition for the interferometer, the frequency difference between the two Raman lasers
can be tuned to be resonant with either the $+k_\mathrm{eff}$ or the $-k_\mathrm{eff}$ transition.

 \begin{figure}[!h]
 \includegraphics[width=8cm]{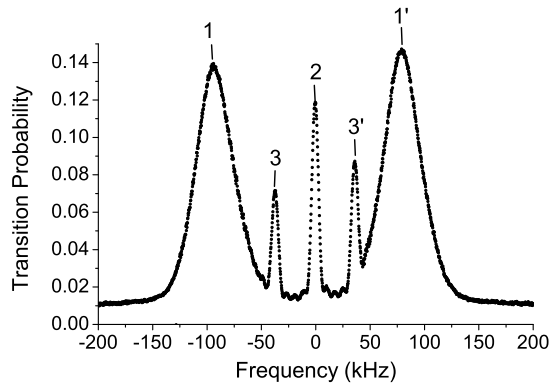}
 \caption{Transition probability as a function of the frequency difference between the two Raman lasers when the atoms have a non-zero velocity in the direction of propagation of the Raman laser beams. The frequency is referenced to the microwave hyperfine transition ($\omega_{0} $). The data have been recorded with laser parameters corresponding to a $\pi$ pulse of 135 $\mu$s duration. Lines (1,1') correspond to the two counter-propagating transitions, line 2 to the copropagating transition between the two $m_F=0$ states and lines 3 to the copropagating magnetic sensitive transitions.}
 \label{Spectre}
 \end{figure}

Figure \ref{Spectre} shows the transition probability as a function of the detuning of the Raman transition with respect to the hyperfine
transition frequency. One can identify the two velocity selective counter-propagation transitions (labelled 1 and 1'), whose widths reflect the velocity distribution of the atomic cloud. In addition to the counter-propagating transitions, we observe transitions due to residual co-propagating Raman coupling, also detuned from resonance by a Doppler shift (lines 2, 3 and 3'). 
When the frequency difference of the Raman lasers is tuned to be resonant with one of the counter-propagating transitions, the second counter-propagating transition and the co-propagating ones induce a light shift (TPLS) on the selected Raman transition used for the interferometer. 


\subsection{Frequency shift due to the 2nd pair of Raman laser beams}

\begin{figure}[!h]
 \includegraphics[width=8.5cm]{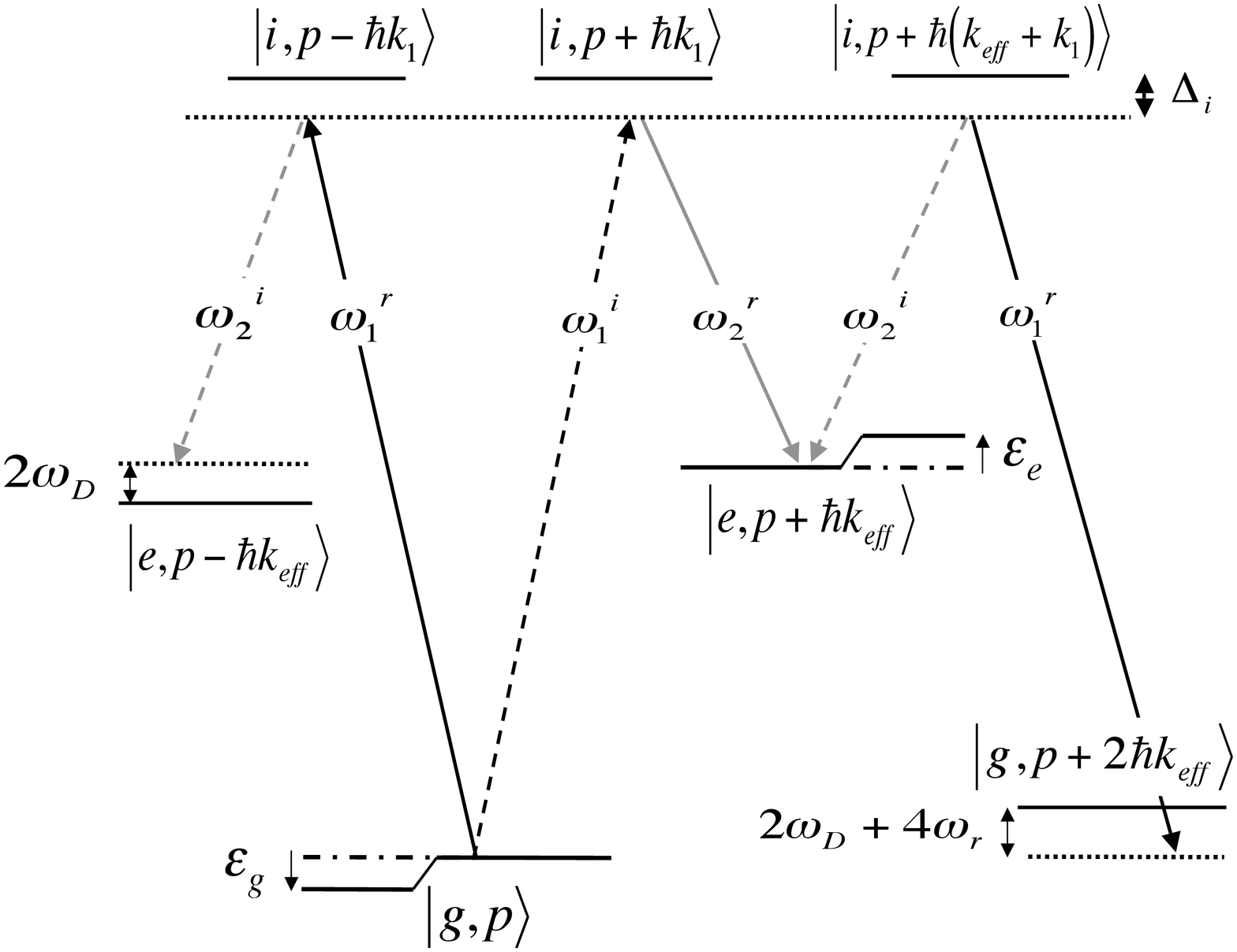}
 \caption{Scheme of the effect on the non-resonant coupling on the two atomic states $|g,\mathbf{p}\rangle$ and $|e,\mathbf{p} + \hbar \mathbf{k}_\mathrm{eff}\rangle$ involved in the atomic interferometer. The two states are coupled together through the selected Raman transition (+\textbf{k} in this particular case). But each state is also coupled to an other one, through an off-resonant Raman transition of opposite wave-vector.}
 \label{Couplages}
 \end{figure}

The TPLS is the differential shift between the two atomic levels corresponding to the atomic states $|g,\mathbf{p}\rangle$ and $|e,\mathbf{p} \pm \hbar \mathbf{k}_\mathrm{eff}\rangle$ involved in the atomic interferometer. The energy of the state $|g,\mathbf{p}\rangle$ is shifted of $\varepsilon_g$ by the off-resonant $|g,\mathbf{p}\rangle \leftrightarrow |e,\mathbf{p} \mp \hbar \mathbf{k}_\mathrm{eff}\rangle$ transition detuned by $\pm2\omega_\mathrm{D}$ (eq.~\ref{epg}), while the energy of the state $|e,\mathbf{p} \pm \hbar \mathbf{k}_\mathrm{eff}\rangle$ is shifted of $\varepsilon_e$ by the off-resonant $|e,\mathbf{p} \pm \hbar \mathbf{k}_\mathrm{eff}\rangle \leftrightarrow |g,\mathbf{p} \mp 2\hbar \mathbf{k}_\mathrm{eff}\rangle$ transition detuned by $\hbar (\mp 2\omega_\mathrm{D}+4 \omega_{r})$ (eq.~\ref{epe}). The two levels are shifted in opposite directions as illustrated in figure~\ref{Couplages}, here for the case of a Raman transition resonant with $+\mathbf{k}_\mathrm{eff}$.

TPLS corrections are calculated from fourth order perturbation
theory. In the case of our system, the level shifts
$\varepsilon_g$ and $\varepsilon_e$ for a $\pm k_\mathrm{eff}$ interferometer are given by:
\begin{eqnarray}
\label{epg}
\varepsilon_g & = & \mp \hbar \frac{\Omega_\mathrm{eff}^2}{4 \;(2\omega_\mathrm{D})}\\
\label{epe}
\varepsilon_e & = & \hbar
\frac{\Omega_\mathrm{eff}^2}{4(\pm2\omega_\mathrm{D}+4\omega_\mathrm{r})},
\end{eqnarray}
where $\Omega_\mathrm{eff}$ is the effective Rabi
frequency corresponding to counter-propagating Raman
transitions. Thus, the shift of the resonance condition depends on the sign of direction of the
selected Raman laser pair (i.e.  $\pm k_\mathrm{eff}$),
quadratically on the Rabi frequency and inversely to the Doppler detuning:

\begin{eqnarray}
\label{nuls2cp}
\begin{array}{ll}
\delta \omega_\mathrm{(TPLS \pm)}& =  \frac{1}{\hbar}(\varepsilon_e-\varepsilon_g)\\[0.3cm]
& =  \frac{\Omega_\mathrm{eff}^2}{\pm 8 \omega_\mathrm{D}}+\frac{\Omega_\mathrm{eff}^2}{4 (\pm 2 \omega_\mathrm{D}+4
\omega_\mathrm{r})}
\end{array}
\end{eqnarray}

\subsubsection{Variation with $\Omega_\mathrm{eff}$}
\begin{figure}[!h]
 \includegraphics[width=8cm]{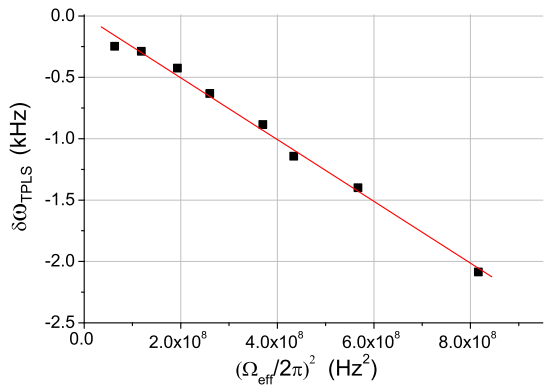}
 \caption{Variation of the frequency shift of the counter-propagating Raman transition versus the square of the Rabi frequency. The Rabi frequency is controlled by changing the Raman laser intensities.}
 \label{TPLS}
 \end{figure}

The frequency shift is measured from fits of the spectrum lines, as displayed in figure \ref{Spectre}, with
different Raman intensities ($\Omega_\mathrm{eff}^2$ is proportional to the product of the intensity of the two lasers). Note that we discriminate it from the shifts independent of
$\mathbf{k}_\mathrm{eff}$, like quadratic Zeeman effect or AC
Stark shift, by alternating measurements $+k_\mathrm{eff}$ and $-k_\mathrm{eff}$, leading to a differential determination of the
effect. The difference in the resonance condition $\Delta
\omega = 2k_\mathrm{eff} v +2\delta \omega_\mathrm{(TPLS)}$ depends
only on the TPLS and the Doppler effect. The Doppler effect does not depend on the Rabi frequency and can be determined by
extrapolating $\Delta \omega$ to $\Omega_\mathrm{eff} = 0$.
The results of these measurements are displayed in figure
\ref{TPLS} as a function of $\Omega_\mathrm{eff}^2$. The
curve clearly shows the quadratic dependence of the
frequency shift with $\Omega_\mathrm{eff}$. For this experimental
configuration the Doppler shift was 85 kHz, and the value of $\delta \omega_\mathrm{(TPLS)}$ for the largest $\Omega_\mathrm{eff}$ ($2\pi\times 27$ kHz at the center of the
beam) is 2.1 kHz in good agreement with the expected 2.4 kHz.

\subsubsection{Position dependance in the Raman laser beams}
\begin{figure}[!h]
 \includegraphics[width=8.5cm]{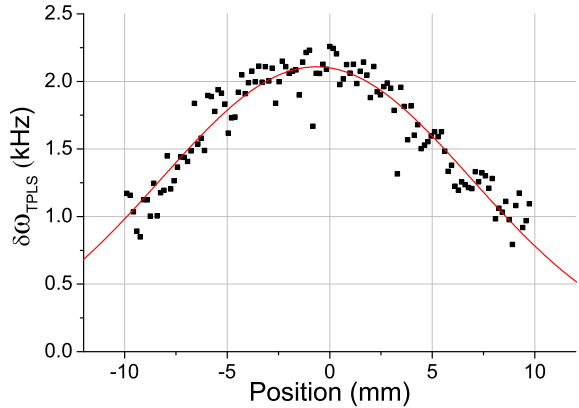}
 \caption{Variation of the two photon light shift with the position in the Gaussian beam of the Raman lasers. The dots correspond to the measurement of the shift and the line to the fit of the data by a Gaussian function.}
 \label{TPLSposition}
 \end{figure}

The laser beams have a Gaussian shape, with a 15 mm
waist (radius at $1/e^2$ in intensity). In order to evaluate the
phase shift, we first measure the TPLS for different atomic positions by
scanning the resonance along their trajectory. As the Raman beam are horizontal, the Doppler shift is constant for the
three pulses and the TPLS varies only with the Raman laser intensity. 
Figure~\ref{TPLSposition} displays the
measured TPLS, proportional to the laser intensity, which
follows exactly the Gaussian profile of the laser
beams.

\subsection{Frequency shift due to the co-propagating transitions}
\label{copro}
In an ideal experiment, with perfect circular polarization and a
Raman detuning $\Delta_i$ large compared to the hyperfine
structure of the intermediate state (201 MHz in the case of the
Caesium atom), co-propagating transitions are forbidden.
In a real experiment, with imperfect polarization and/or finite Raman detuning, co-propagating transitions
are slightly allowed, and will lead to additional TPLS.

\subsubsection{TPLS induced by $m_F=0$ to $m_F=0$ co-propagating transitions}

Imperfect polarization leads to a residual combination of
$\sigma^+_1 / \sigma^{+}_{2}$ and $\sigma^{-}_{1} /
\sigma^{-}_{2}$ in the co-propagating beams and allows coupling
between $m_F = 0$ states. As the momentum exchanged $\hbar
\mathbf{k}_\mathrm{eff} = \hbar(\mathbf{k_1}-\mathbf{k_2}) \simeq 0$, the
Doppler and recoil effect are negligible and the resonance
condition is $ \Delta \omega_\mathrm{laser} = \omega_{0}$ (line 2 of
figure \ref{Spectre}). The Rabi frequency corresponding
to the transition $\Delta m_F = 0$ is determined experimentally using the residual co-propagating transition
probability, $P = 0.1$  at full Raman laser power. It gives
$\frac{\Omega_\mathrm{eff}}{\Omega_{00}} = 5$, which can be explained
by an error of linear polarization of one of the Raman laser of
2\% in power. The detuning of this transition, compared to the two
counter-propagating transitions, depends on the Doppler
and recoil shift:
\begin{equation}
\delta \omega_\mathrm{(TPLS \pm)}^{00} \simeq \frac{1}{4}\frac{\Omega_{00}^2}{\pm \omega_\mathrm{D} + \omega_\mathrm{r}}
\end{equation}
For $\omega_\mathrm{D} = 2 \pi\times85$ kHz and $\Omega_\mathrm{eff} \simeq 2\pi\times 27 $ kHz we
find an effect due to this coupling smaller than 100 Hz.

\subsubsection{TPLS induced by $m_F=0$ to $m_F=\pm2$ co-propagating transitions}

The second source of residual co-propagating transitions stems from the coupling of $|F,m_F=0\rangle \leftrightarrow |F',m_{F'}=\pm 2\rangle$ by the co-propagating
Raman laser pairs ($\sigma^{+}_1$,$\sigma^{-}_2$) and ($\sigma^{-}_1$,$\sigma^{+}_2$). Because of the hyperfine splitting in the intermediate state, there are two paths for the Raman transition. Both
transitions interfere destructively when the detuning compared to the intermediate state $\Delta_i$ is larger than the hyperfine splitting of the intermediate state, and so in this case the
transition strength is zero. However, in our experimental set up, with $\Delta_i \simeq 2 \pi\times780$ MHz and the $\Delta_{HFS} = 2 \pi\times201$ MHz the ratio between the Rabi frequency
of counter-propagating transitions $\Omega_\mathrm{eff}$ and the Rabi frequency of the co-propagating $\Delta m_F = \pm 2$ transition $\Omega_{02}$ is 6.1, leading to a transition
probability of 6.4 $\%$ in good agreement with the experimental value (see figure~\ref{Spectre}). These transition resonance conditions depend on the magnetic field amplitude as $\Delta \omega_\mathrm{laser} = \omega_0 \pm 2 \alpha B$, where $\alpha = 2\pi\times350$ kHz/G for Caesium. With a calculation similar to the one used to obtain eq.\ref{nuls2cp}, we deduce the two photon light shifts $\delta \omega _\mathrm{(TPLS \pm)}^{02}$ induced
by the magnetically sensitive transitions for the $\pm k_\mathrm{eff}$ case to be:

\begin{eqnarray}
\label{nuls2B}
\begin{array}{c}
\delta \omega_\mathrm{(TPLS \pm)}^{02} = \frac{\Omega_{02}^2}{4} \left(\frac{1}{\pm
\omega_\mathrm{D}+\omega_\mathrm{r}+2\alpha B}+\frac{1}{\pm
\omega_\mathrm{D}+\omega_\mathrm{r}-2\alpha B}\right)
\end{array}
\end{eqnarray}

\begin{figure}[!h]
 \includegraphics[width=8.5cm]{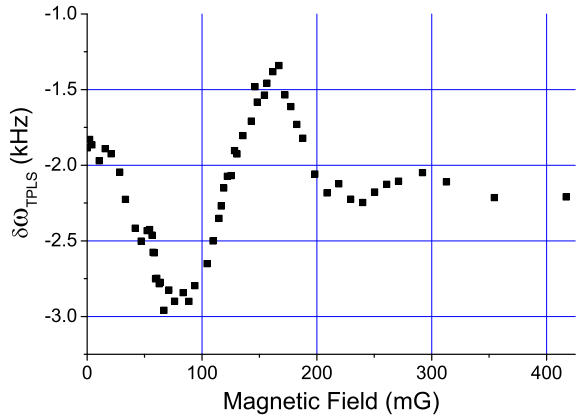}
 \caption{Variation of the two photon light shift versus the bias magnetic field. A resonance  appears when the Zeeman shift equals the Doppler shift. }
 \label{TPLS_B}
 \end{figure}

The first term in eq. \ref{nuls2B} is due to the coupling with $|F',m_{F'}=+ 2\rangle$ whereas the second term is induced by the coupling with $|F',m_{F'}=- 2\rangle$. It
is clear from eq. \ref{nuls2B} that a residual magnetic contribution appears in the half difference and creates a magnetic sensitivity in addition
to the standard quadratic Zeeman effect. This contribution to the two photon light shift is measured by changing the bias field in
the Raman interaction zone. Using the differential method previously described, we show in figure \ref{TPLS_B} the variation
of the total TPLS with the bias field. The resonance around $130$ mG corresponds to the case where a magnetic co-propagating transition and the
counter-propagating transition are resonant simultaneously. Previous measurements have been performed with a 31 mG magnetic field bias.

\section{Impact on the interferometer phase shift}

\subsection{Theoretical derivation of the phase shift}

In the following we will consider interferometers
constituted of three Raman pulses in a $\pi/2-\pi-\pi/2$
sequence. If the TPLS is constant during the interferometer it is
equivalent to a fixed frequency shift of the Raman
transition. In that case, It is well known that no phase shift is introduced in the interferometer.
On the contrary, a fluctuation of the TPLS during the
interferometer sequence leads to a phase shift given by:
\begin{equation}
\Phi_\mathrm{TPLS}=\int_{-\infty}^{+\infty}g(t)\delta \omega_\mathrm{TPLS}(t)dt.
\end{equation}
where $g(t)$ is the sensitivity function of the atom interferometer, defined in \cite{IEEE}.

\subsubsection{Case  $\Omega_\mathrm{eff}\tau=\pi/2$ }

In the case where the interaction pulses are short  enough that
one can neglect the variation of the TPLS during the
pulses, and that the area of first and
last pulse fulfils the $\Omega_\mathrm{eff}\tau=\pi/2$ condition,
the two-photon interferometer phase shift can be approximated by:

\begin{equation}
\label{TPPS} \Phi_\mathrm{TPLS}=\left(\frac{\delta
\omega^{(1)}_\mathrm{TPLS}}{\Omega^{(1)}_\mathrm{eff}}-\frac{\delta
\omega^{(3)}_\mathrm{TPLS}}{\Omega^{(3)}_\mathrm{eff}}\right),
\end{equation}
where $\omega^{(i)}_\mathrm{TPLS}$ and $\Omega^{(i)}_\mathrm{eff}$ are the TPLS
and the Rabi frequencies of the i-th pulse, respectively.
One might notice that the frequency shift during the $\pi$ pulse
does not contribute to the interferometer phase shift. Moreover, as all
components of the TPLS, counter-propagating and
co-propagating terms, increase as the square of the Rabi
frequencies, the interferometer phase shift scales linearly with the
Raman laser power.

In the limit where the co-propagating
transitions are negligible (perfect polarization and very large
Raman detuning) and the dominant source of TPLS is due to the
counter-propagating transition, the phase shift
of the interferometer can be simplified as:

\begin{equation}
\label{TPPS2}
\Phi^\mathrm{counter}_\mathrm{TPLS}=\left(\frac{\Omega^{(1)}_\mathrm{eff}}{4\delta_D^{(1)}}-\frac{\Omega^{(3)}_\mathrm{eff}}{4\delta_D^{(3)}}\right),
\end{equation}
where $\delta_D^i$ is the Doppler shift for the i-th
pulse.

\subsubsection{Case  $\Omega_\mathrm{eff}\tau\neq\pi/2$ }
   
More generally, the interferometer phase shift can be calculated when the $\pi/2$ pulse condition is no longer fulfilled. This appears when the Rabi frequency drifts
due to changes in power or polarization of the Raman lasers. Generalising the formalism of the sensitivity function to the case where $\Omega_\mathrm{eff}\tau\neq\pi/2$ allows deriving the interferometer phase shift:
\begin{eqnarray}
\label{TPPSgeneral}
\begin{array}{lll}
\Phi_\mathrm{TPLS}& = &\frac{\delta \omega^{(1)}_\mathrm{TPLS}}{\Omega^{(1)}_\mathrm{eff}}\tan\left(\frac{\Omega^{(1)}_\mathrm{eff}\tau^{(1)}}{2}\right)\\[0.3cm]
                         &&-\frac{\delta\omega^{(3)}_\mathrm{TPLS}}{\Omega^{(3)}_\mathrm{eff}}\tan\left(\frac{\Omega^{(3)}_\mathrm{eff}\tau^{(3)}}{2}\right)
\end{array}
\end{eqnarray}
Usually, the Rabi frequencies and pulse durations can be
taken equal for the first and the last pulses; the expression of
the interferometer shift is then:
\begin{equation}
\label{TPPSgeneral2}
\Phi_\mathrm{TPLS}=\frac{\left(\delta \omega^{(1)}_\mathrm{TPLS}-\delta\omega^{(3)}_\mathrm{TPLS}\right)}{\Omega_\mathrm{eff}}\tan\left(\frac{\Omega_\mathrm{eff}\tau}{2}\right)\
\end{equation}
As before, for a dominant counter-propagating
transition, the previous expression can be simplified to:
\begin{equation}
\label{TPPSgeneral3}
\Phi^\mathrm{counter}_\mathrm{TPLS}=\Omega_\mathrm{eff}\left(\frac{1}{4\delta_D^{(1)}}-\frac{1}{4\delta_D^{(3)}}\right)\tan\left(\frac{\Omega_\mathrm{eff}\tau}{2}\right)\
\end{equation}

An other aspect of the influence of the TPLS on the
atomic phase shift concerns the stability of the experiment versus
the experimental parameters fluctuations, in particular the Raman
laser power. As Rabi frequency fluctuations are small in
relative values (typically smaller than 10 \%), we can develop eq. \ref{TPPSgeneral3} to first order
in $\delta\Omega_\mathrm{eff}$ close to the usual conditions
$\Omega_\mathrm{eff}\tau=\pi/2$ and find:

\begin{equation}
\label{TPPSlimite}
\delta\Phi_\mathrm{TPLS}=\left(1+\frac{\pi}{2}\right)\frac{\delta\Omega_\mathrm{eff}}{\Omega_\mathrm{eff}}\Phi_\mathrm{TPLS}
\end{equation}
Similar calculations may be derived to extract the
dependance on the duration of the pulse or the
Doppler detuning. As the stability of the latter
parameters is much better controlled in cold atom interferometers,
no measurable influence on the short term stability of the
interferometer is expected.

\subsection{Case of a gravimeter}
We first consider the case of the gravimeter developed at
SYRTE and described in detail in~\cite{GraviAPB1,GraviAPB2}. In
this compact experimental set-up, cold $^{87}Rb$ atoms are trapped in a 3D MOT in 60 ms and
further cooled during a brief optical molasses phase before being
released by switching off the cooling lasers. During their free
fall over a few centimetres, the interferometer is created by driving the
Raman laser in the vertical direction, with pulses separated by
free evolution times of $T=50$ ms. The first Raman pulse occurs
17 ms after releasing the atoms. The Doppler shift at the
first pulse is thus relatively small, about 400 kHz, and gets
large, about 3 MHz, for the last pulse. With a Rabi frequency of
40 kHz, this leads to a TPLS of about 22 mrad for
counter-propagating transition. Following the method of paragraph \ref{copro}, we find a 11 mrad for
co-propagating ones. This corresponds to a large shift for
the gravity measurement of about $8.10^{-8}g$.

\begin{figure}[!h]
 \includegraphics[width=8.5cm]{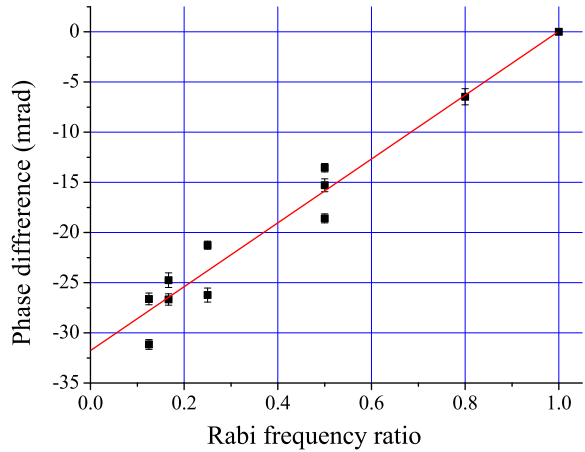}
 \caption{Variation of the gravimeter phase shift due to the two photon light shift versus the Rabi frequency ratio, keeping the pulse area constant. To remove drifts from other sources, the phase shift is measured in a differential way by changing the Rabi frequency up to 40 kHz, which is the maximum available. The dotted line is a linear fit of the shifts.}
 \label{ShiftGravi}
 \end{figure}

To measure the bias on the atomic interferometer phase due to
TPLS, we exploit its dependence with the Rabi frequency. The
principle of this measurement is based  on a differential method,
where one performs an alternating sequence of
measurements of the interferometer phase with two different Rabi
frequencies $\Omega_\mathrm{eff}$ and $\Omega'_\mathrm{eff}$, but keeping the
areas of the pulses constant by changing the duration of the
pulses $\tau$. The Rabi frequency is modulated
with the power of the Raman lasers. In practice, the
differential measurement is performed by alternating sequences of
measurements with four different configurations : ($\Omega_\mathrm{eff},
+k_\mathrm{eff}$), ($\Omega_\mathrm{eff}, - k_\mathrm{eff}$), ($\Omega'_\mathrm{eff}, +
k_\mathrm{eff}$) and ($\Omega'_\mathrm{eff}, - k_\mathrm{eff}$). After averaging for 5 minutes, we extract
the difference of the TPLS between the two Rabi
frequencies with an uncertainty below 1 mrad.
This measurement was repeated for various $\Omega'_\mathrm{eff}$, keeping
$\Omega_\mathrm{eff}$ fixed. The phase differences are displayed in
figure~\ref{ShiftGravi} in according to the ratio of the Rabi
frequency $\Omega'_\mathrm{eff}/\Omega_\mathrm{eff}$. The results
clearly demonstrate the linear dependence of the phase shift with the
Rabi frequency. The fit of the data allows to extract a 32 mrad shift, in very good agreement with the
expected value (33 mrad), deduced from eq.~\ref{TPPS}. Deviations from the linear behavior and
discrepancies (up to $\pm 10 \%$) between different measurements
that correspond to the same ratio cannot be explained by
uncontrolled fluctuations in the Rabi frequencies, as the
simultaneous monitoring of the laser intensities showed stability
at the per cent level during the course of the measurements. We
demonstrated that these fluctuations were correlated with changes
of the polarization of the Raman beams, which modulate the
contribution to the phase shift of the undesired co-propagating
transitions.

\begin{figure}[!h]
 \includegraphics[width=8.5cm]{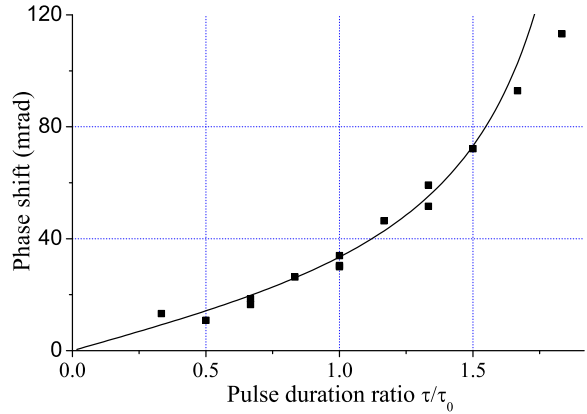}
 \caption{Variation of the gravimeter phase due to the two photon light shift versus the pulse duration ratio, keeping the the Rabi frequency constant (40 kHz). To remove drifts from other sources, the phase shift is measured in a differential way with respect to the optimum pulse duration (6 $\mu$s). The line corresponds to the calculated shift.}
 \label{shiftVsTau}
 \end{figure}
We perform a complementary measurement by changing  the
duration of the first and last pulse simultaneously, while keeping $\Omega_\mathrm{eff}$ constant. The
acquisition is performed with a similar differential method than
previously, but now alternating between different pulse
durations and a pulse duration of 6 $\mu s$,
when the $\pi/2$ pulse condition is fulfilled. The data are then
shifted by the bias deduced from previous measurements
with $\tau=6$ $\mu s$ (result of figure~\ref{ShiftGravi}) and
displayed in figure~\ref{shiftVsTau} . Using
eq.~\ref{TPPSgeneral2}, the fit of the data gives a Rabi frequency
of 39 kHz , in very good agreement with the expected value (42
kHz). A small deviation appears for long pulse durations when the
areas of each of the two pulses is close to $\pi$.

In the case of our gravimeter, where cold atoms are
dropped from rest, the TPLS is very large, almost two orders of
magnitude above the pursued accuracy. In principle, this effect
can be measured accurately by alternating measurements with
different Rabi frequencies. But, it seems desirable to decrease
the effect by operating with lower Rabi frequencies, the drawback
being an increased velocity selectivity of the Raman pulses. A
more stringent velocity selection, or smaller temperatures, are
then required in order to preserve a good fringe contrast. In the case of
a fountain gravimeter, where the atoms are launched upwards at a
few m/s, the Doppler shift at the first and last
pulse is much larger, considerably reducing this
effect. For the parameters of the Stanford gravimeter
\cite{Ackim} (long pulse duration of 80 $\mu$s and time between
pulses of 160 ms), we find a phase shift of 0.8 mrad, which
corresponds to $2.10^{-10}g$.

\subsection{Case of a gyroscope-accelerometer}

In the case of the gyro-accelerometer, the mean velocity of the wave-packet is not collinear with the effective Raman wave vector. Consequently, the atomic phase shift measured with the
interferometer is sensitive to the rotation rate \textbf{$\Omega$} in addition to the acceleration \textbf{a}. As in the case of the gravimeter, the atomic phase is shifted by the TPLS and by other systematics (labelled $\Phi_0$), for instance the phase shift induced by the laser wave-front distortions~\cite{jerome}. The total phase shift is
expressed by :

\begin{equation}
\Phi = \mathbf{k}_\mathrm{eff} \cdot \mathbf{a} T^2 + 2
\mathbf{k}_\mathrm{eff} \cdot
\mathbf{V}\times\mathbf{\Omega} T^2 + \Phi_0 +
\Phi_\mathrm{TPLS}
\end{equation}

Our gyroscope-accelerometer uses a double interferometer with two atomic clouds following the same trajectory but with opposite directions to discriminate between acceleration and rotation phase shifts. Moreover, our experiment is designed to measure different axis of rotation and acceleration according to direction of propagation of the Raman laser beams. We will illustrate the impact of the TPLS on the interferometer in the configuration where the Raman lasers are directed along the vertical direction. 

\begin{figure}[!h]
 \includegraphics[width=8.5cm]{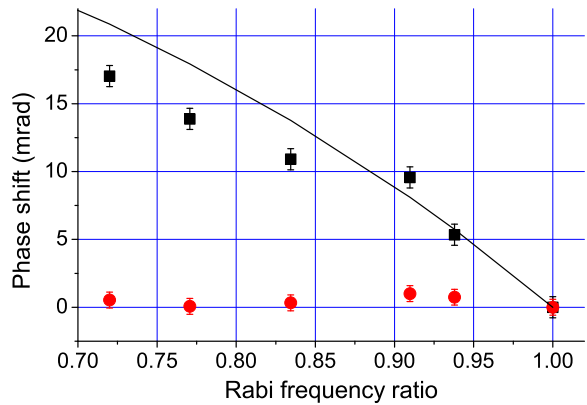}
 \caption{Variation of the rotation and acceleration phases (resp. circles and squares) due to the two photon light shift versus the Rabi frequency ratio, keeping the pulse duration constant (7.6 $\mu$s). Acquisitions have been recorded for a total interaction time between the first and the last pulse of 40 ms. To remove drifts from other sources, the phase shift is measured in a differential way with respect to the maximum Rabi frequency available at the location of the $\pi$/2 pulses (33 kHz). The line shows the calculated shift for acceleration signal.}
 \label{ShiftGyro40}
 \end{figure}

The measurement is realized in the same way than for the gravimeter experiment by comparing atomic phases with high and low Rabi frequency, changing the Raman laser power but keeping the pulse duration constant to 7.6~$\mu$s. In order to enhance the TPLS signal we decrease the time between pulses to 20 ms. Indeed, the Doppler effect is reduced and the available laser power on the side of the Gaussian laser profile is increased. The first pulse occurs 15 ms before the apogee and the third 25 ms after the apogee, corresponding respectively to a Doppler shift of about $\omega_\mathrm{D}^{(1)}= 2\pi\times$344~kHz and $\omega_\mathrm{D}^{(3)}= 2\pi\times$574~kHz. The Rabi frequency for each pulse is approximately 33~kHz, then the TPLS expected from eq.\ref{TPPS} is about 38 mrad for each interferometer. As this shift is similar for the two interferometers when the two atomic clouds perfectly overlap and experiment the same TPLS, it bias the acceleration signal only. Figure~\ref{ShiftGyro40} displays the variation of the acceleration and rotation signals with the Rabi frequency. Acceleration shift (squares) varies in a good agreement with expected shift (continuous curve) calculated from eq.~\ref{TPPSgeneral}. Rotation shift (circles) shows no dependance on the Rabi frequency and illustrates that the rejection from the acceleration signal is efficient. Nevertheless, fluctuations from expected behaviour are clearly resolved and repeatable. We attribute these deviations to wave-front distortions of the Raman laser beams. Indeed, when the atomic trajectories do no perfectly overlap, a residual bias appears on the rotation due to unperfected cancellation. This bias depends on the details of the wave-front distortions weighted by the actual atomic cloud distributions, and is modified when the Rabi frequency is changed. In the usual conditions, with interrogation time of 80~ms, the acceleration shift is reduced to about 12 mrad thanks to the increase of the Doppler shift and reduction of the Rabi frequency on the side of the Gaussian Raman beams.

We finally estimate the impact of the Raman laser power fluctuations on the stability of the rotation signal in usual conditions (interrogation time of 80 ms).  We performed a complementary measurement by recording the interferometer signal and the Raman laser power in the same time when applying a modulation of the laser powers of 10$\%$. The modulation is applied by attenuating the radio-frequency signal sends to the acousto-optic modulator used to generate the Raman pulses. The power of the lasers was recorded during the third pulse thanks to a photodiode measuring the intensity on the edge of the laser beams. We found a small dependance of the rotation signal on the power fluctuation of $6.10^{-9}$ rad.s$^{-1}$/$\%$, which can limit the long term stability of the gyroscope. As no dependance was expected, we attribute it again to non perfect superimposition of the two atomic clouds trajectories, leading to a different Rabi frequencies experienced by the two clouds.

\section{Conclusion}
We have shown that the use of retro-reflected Raman lasers in atom interferometers induces off-resonant Raman transitions, which have to be taken into account in order to achieve best accuracy and stability of interferometers. We have first quantitatively evaluated the effect on the resonance condition for each off-resonant line: the other counter-propagating transition, which can not be avoided in the retro-reflected design, and co-propagating transitions arising mainly from imperfections in the polarisation.
Then, we have measured the impact of this two photon light shift on the phase of two atom interferometers: a gravimeter and a gyroscope. In particular, we show that this shift is an important source of systematic errors for acceleration measurements. Nevertheless, it can be measured accurately by modulating the Raman laser power and/or the pulse durations. Our study has also shown that it can impact the stability of the two sensors if the polarization and/or the power of the Raman lasers fluctuate.

The TPLS appears as a drawback of using retro-reflected Raman laser beams. But, as it can be well controlled, it does not reduce the benefit from this geometry, whose key advantage is to drastically limit the bias due to wave-front aberrations, which is larger and more difficult to extrapolate to zero.

This study can be extended to other two photon transitions (like Bragg transitions) and to other possible polarization configurations (when using linear instead of circular polarizations), when using the Doppler effect to select the transition. If the signal is generated from the subtraction of the phase shifts of two independent atomic clouds, e.g. gradiometers or gyroscopes, perfect common mode rejection,
is required to suppress this effect. In our case this means a perfect overlap of atomic trajectories.

Finally, the two photon light shift can be drastically reduced by increasing the Doppler effect and/or using colder atoms, allowing to reduce the Rabi frequency during the Raman pulses. By contrast, for set-up with intrinsic small Doppler effect, as for space applications~\cite{varenna,space}, this effect becomes extremely large and has to be taken into account in the design of the experiment.

\begin{acknowledgments}
We express our gratitude to F. Biraben for pointing out this effect and to P. Cheinet for earlier contributions to the gravimeter experiment. We would like to thank the Institut Francilien pour la Recherche sur les Atomes Froids (IFRAF), the European Union (FINAQS contract), the D\'{e}l\'{e}gation G\'{e}n\'{e}rale pour l'Armement (DGA) and the Centre National d'\'{E}tudes Spatiales (CNES) for financial supports. B. C., J. L.G. and T. L. thank DGA for supporting their work. 
\end{acknowledgments}


\end{document}